\documentclass[12pt]{article}
\usepackage{epsf}

\newcommand{\be}{\begin{equation}}
\newcommand{\ee}{\end{equation}}

\title{QCD string model for hybrid adiabatic potentials}
\author{Yu.S.Kalashnikova\thanks{e-mail: yulia@vxitep.itep.ru}  and
D.S.Kuzmenko\thanks{e-mail: kuzmenko@vxitep.itep.ru}}
\date{\it Institute of Theoretical and Experimental Physics,\\
117218, B.Cheremushkinskaya 25, Moscow, Russia}

\begin{document}

\maketitle

\begin{abstract}
Hybrid adiabatic potentials are considered in the framework of
the QCD string model. The einbein field formalism is applied to 
obtain the large-distance behaviour of adiabatic potentials.
The calculated excitation curves are shown to be the result of 
interplay between potential-type longitudinal and string-type 
transverse vibrations. The results are compared with recent
lattice data.

\end{abstract}
\date{}
\maketitle

\section{Introduction}

There is no doubts now that hadrons with explicit gluonic degrees 
of freedom should exist. This idea is supported not only by
general arguments from QCD, but also by lattice simulations of
pure Yang-Mills theory. In the absence of exact analytical methods
of nonperturbative QCD one relies upon models to describe gluonic
mesons, so that the challenging question arises of how to
introduce effective degrees of freedom for soft (constituent) glue.

There is a lot of indications now that gluonic mesons are
already found experimentally, but the conclusive evidences have never
been presented; there is no hope that in the nearest future
data analyses could shed light on this problem and to offer necessary
feedback for model building. The current situation is
such that the predictions of different models on hadronic spectra
and decays are involved in order to pin-point the signatures for
gluonic mesons.

On the other hand, lattice calculations are now accurate enough to 
provide reliable data on constituent glue and to check model predictions.
In this regard recent measurements of gluelump \cite{gluelump} and
hybrid adiabatic potentials \cite{lattice} are of particular interest. 
These simulations measure the spectrum of the glue in the presence of
infinitely heavy adjoint source (gluelump) and in the presence of
static quark and antiquark separated by some distance $R$. These systems
are the simplest ones and play the role of hydrogen atom of soft glue
studies, as the complicated problem of the centre-of-mass motion 
separation is not relevant here. 

Hybrid adiabatic potentials enter heavy hybrid mass estimations in the
Born-Oppenheimer approximation: these potentials are to be inserted 
into $Q \bar Q$ Schroedinger equation in order to obtain spectra of 
hybrids with heavy quarks. The large $R$ limit is interesting \begin{it}
per se, \end{it} as the formation of confining string is expected at large
distances, so the direct measurements of the string fluctuations become
available and the possibility exists to discriminate between different
models of the effective string degrees of freedom.

\section{Constituent gluons at the end of the string}

The notion of confinement is usually described in terms of area law
asymptotics for the Wilson loop expectation value, defined as an
integral along some closed contour $C$, averaged over gluonic
vacuum configurations:
\be
\langle W(C)\rangle = \mathrm{Tr} \langle
P\exp{\;ig\oint_CA_{\mu}dz_{\mu}}\rangle, \label{WL}
\ee
where trace is taken over colour indices. The area law asymptotics implies
that 
\be
\langle W(C)\rangle \rightarrow N_C\exp(-\sigma S),
\label{area}
\ee
where $N_C$ is the number of colours, $\sigma$ is the string tension, and
$S$ is the surface bound by the closed contour $C$. As the initial
expression (\ref{WL}) depends only on the contour, the area in
(\ref{area}) should depend only on the contour too, and should be the
minimal area. 

The area law asymptotics provides the action of the string, and in the
case of minimal area this string is also "minimal". The effective string
model should be arranged to allow the extra degrees of freedom to populate
the string and to be responsible for more complicated string
configurations. In what follows these extra degrees of freedom are defined
in the framework of the QCD string model. This model deals with quarks and
point-like gluons propagating in the confining QCD vacuum, and is based on
Vacuum Background Correlators method \cite{Lisbon}. 

The QCD string model for gluons is derived from the perturbation theory 
in the nonperturbative confining background developed in \cite{Pert}.
The main idea is to split the gauge field as
\be
A_{\mu}=B_{\mu}+a_{\mu},
\label{split}
\ee
which allows to distinguish clearly between confining gluonic field
configurations $B_{\mu}$ and confined valence gluons $a_{\mu}$. Confining
QCD vacuum is given by the set of gauge invariant field strength
correlators made of $B_{\mu}$, which are responsible for the area law
asymptotics (\ref{area}), while the valence gluons are treated as
perturbation at this confining background.

The starting point is the Green function for the gluon propagating in the
given external field $B_{\mu}$ \cite{Pert}:
\be
G_{\mu\nu} (x,y) = (D^2(B)\delta_{\mu\nu} + 2igF_{\mu\nu}(B))^{-1}.
\label{Green}
\ee
The term, proportional to $F_{\mu\nu}(B)$ is responsible for the gluon
spin interaction, it can be treated as perturbation \cite{S}, and we 
neglect it for a moment. The next step is to use Feynman-Schwinger
representation for the quark-antiquark-gluon Green function, 
which, for the case of static quark and antiquark, is reduced to 
the form
\be
G(x_g, y_g) = \int ds \int Dz_g \exp (-K_g) \langle
{\cal W}\rangle_B,
\label{FS}
\ee
where angular brackets mean averaging over background field. The
quantity $K_g$ is the kinetic energy of gluon (to be specified below),
and all the dependence on the vacuum background field is contained in
the generalized Wilson loop $\cal W$, depicted in Fig.1.

\begin{figure}[t!]
\epsfxsize=7cm
\centering
\epsfbox{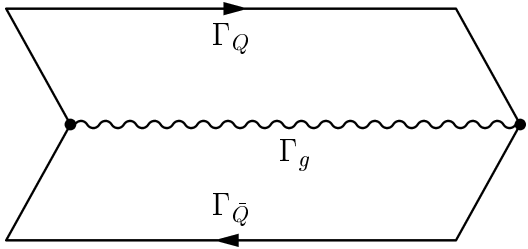}
\caption{Hybrid Wilson loop.}
\end{figure}

The main assumption of the QCD string model is the minimal area law, 
which yields for the configuration $\cal W$ the form \cite{hybrids}
\be
\langle {\cal W}\rangle_B = \frac{N_C^2-1}{2}\exp (-\sigma(S_1+S_2)),
\label{W}
\ee
where $S_1$ and $S_2$ are the minimal areas inside the contours formed by
quark and gluon and antiquark and gluon trajectories correspondingly.

\section{Einbein field form of the gluonic Lagrangian}

To define the form of gluon kinetic energy we note that
the action of a particle in the external vector field is invariant
under reparametrization transformations, and, of course, it remains true
after averaging over the background. So, to proceed further we are to fix
the gauge, and the most natural way to do this is to identify the
proper time $\tau$ of the path integral representation with
the physical time $x_g^{0}$. Then the action of the system can be
immediately read out of the representation (\ref{FS}):
$$
A=\int d\tau \left \{ -\frac{\mu}{2}+\frac{\mu\dot r^2}{2} - \sigma
\int^1_0 d\beta_1
\sqrt{(\dot w_1 w_1')^2-\dot w_1^2w^{'2}_1}-\right.
$$
\be
\left.-\sigma\int^1_0 d\beta_2
\sqrt{(\dot w_2 w_2')^2-\dot w_2^2w^{'2}_2}\right \},
\label{Lagr}
\ee
where the minimal surfaces $S_1$ and $S_2$ are parametrized by the
coordinates $w_{i\mu}(\tau,\beta_i) , ~i=1,2,~
\dot w_{i\mu}= \frac{\partial w_{i\mu}}{\partial \tau},~
w'_{i\mu}= \frac{\partial w_{i\mu}}{\partial \beta_i}.$
In what follows the straight-line ansatz is chosen for the
minimal surface:
\be
w_{i0}=\tau,~~~\vec w_{1,2} = \pm (1-\beta) \frac{\vec R}{2}+\beta
\vec r.
\label{surf}
\ee
The quantity $\mu=\mu(\tau)$ in the expression (\ref{Lagr}) is the 
so-called einbein field \cite{einbein}; here one is forced to
introduce it, as it is the only way to obtain the meaningful dynamics for
the massless particle. 

Let us introduce another set of einbein
fields,
$\nu_i=\nu_i(\tau,\beta_i)$ to get rid of Nambu-Goto square roots in
(\ref{Lagr}) \cite{DSK}. The resulting Lagrangian takes the form
$$
L=-\frac{\mu}{2}+\frac{\mu\dot r^2}{2}-
\int^1_0 d\beta_1\frac{\sigma^2 r^2_1}{2\nu_1}-\int^1_0
d\beta_1\frac{\nu_1}{2}(1-\beta_1^2 l_1^2)-
$$
$$
-\int^1_0 d\beta_2\frac{\sigma^2 r^2_2}{2\nu_2}-\int^1_0
d\beta_2\frac{\nu_2}{2}(1-\beta_2^2 l_2^2),
$$
\be
l^2_{1,2}=\dot r^2-\frac{1}{r^2_{1,2}}(\vec r_{1,2}\dot{\vec r})^2,~~
\vec r_{1,2}=\vec r\pm \frac{\vec R}{2},
\ee
and the corresponding Hamiltonian reads
\be
H=H_0+\frac{\mu}{2}
+\int^1_0 d\beta_1\frac{\sigma^2 r^2_1}{2\nu_1}
+\int^1_0
d\beta_2\frac{\sigma^2 r^2_2}{\nu_2}+\int^1_0
d\beta_1\frac{\nu}{2}+
\int^1_0
d\beta_2\frac{\nu_2}{2},
\ee

$$
H_0=\frac{p^2}{2(\mu+J_1+J_2)}+
$$
$$
\frac{1}{2\Delta (\mu+J_1+J_2)}
\left\{ \frac{(\vec p\vec r_1)^2}{r^2_1}J_1(\mu+J_1)+
\frac{(\vec p \vec r_2)^2}{r^2_2}J_2(\mu+J_2)+\right.
$$
\be
\left.\frac{2J_1J_2}{r^2_1r^2_2}(\vec r_1\vec r_2)(\vec p\vec r_1)
(\vec p\vec r_2)\right\},
\label{Ham}
\ee

$$
\Delta=(\mu +J_1)(\mu +J_2)-
J_1J_2\frac{(\vec r_1\vec r_2)^2}{r^2_1r^2_2},~~
J_i=\int^1_0 d\beta_i\beta_i^2\nu_i(\beta_i),~~i=1,2.
$$

At first glance, the Hamiltonian (\ref{Ham}) looks tractable. Clearly, 
quantities  $\mu$ and $\nu_i$ play the role of gluon constituent mass and 
energy density distributions along the string respectively. Nevertheless,
introducing einbeins does not do miracles for us. These redundant
variables are to be found from the conditions
\be
\frac{\partial H}{\partial\mu}=0,
~~\frac{\delta H}{\delta \nu_i(\beta_i)}=0,
\label{constraints}
\ee
as the equations (\ref{constraints}) play the role of second class
constraints. One should do it before quantization and substitute
 the resulting values into the Hamiltonian. Such procedure
is hardly possible analytically even at the classical level, and after
quantization these extremal values of einbeins would become nonlinear
operator functions of coordinates and momenta with inevitable ordering
problems arising. 

In what follows we use the approximate einbein field method, which treats
the einbeins as $c$-number variational parameters. The eigenvalues
of the Hamiltonian (\ref{Ham}) are found as functions of $\mu$ and $\nu_i$
and minimized with respect to eibeins to obtain the physical spectrum.
Such procedure works surprisingly well in the QCD string model
calculations, with the accuracy of about 5-10\% \cite{rotstring}.

\section{Dynamical regimes of the gluonic Hamiltonian}

Even with the simplifying assumptions of the approximate einbein field
method the problem remains complicated due to the presence of the terms
$J_i$ responsible for the string inertia. If one neglect these terms, then
the einbeins are eliminated explicitly from the Hamiltonian (\ref{Ham}),
and one arrives at the potential model Hamiltonian
\be
H=\sqrt{p^2}+ \sigma r_1 + \sigma r_2.
\label{pot}
\ee 
It appears, however, that the neglect of string inertia is justified only
for $R \leq 1/\sqrt{\sigma}$ \cite{vibr}. Indeed, in the einbein field
method the
potential regime corresponds to the case of $\nu_i$ independent of
$\beta_i$. For example, for $R \ll 1/\sqrt{\sigma}$ one has
\be
E_n(R) = 2^{3/2}\sigma^{1/2}
(n+3/2)^{1/2}+\frac{\sigma^{3/2}R^2}{2^{3/2}(n+3/2)^{1/2}},
\label{ES}
\ee
$$
\mu_n(R) = 2^{1/2}\sigma^{1/2}
(n+3/2)^{1/2}-\frac{\sigma^{3/2}R^2}
{2^{5/2}(n+3/2)^{1/2}},
$$
$$
\nu_{1,2 n}(R) = \frac{(n+3/2)^{1/2}\sigma^{1/2}}{2^{1/2}}
+\frac{3\sigma^{3/2}R^2}{2^{7/2}(n+3/2)^{1/2}},
$$ 
where $n$ is the number of oscillator quanta. The last line in (\ref{ES})
readily gives $J_{1,2}/\mu\approx 1/6$. The situation here is similar
to the one in the light quark, glueball and gluelump sectors: the
corrections due to the string inertia are sizeable, but not large, and 
can be taken into account as perturbation \cite{S}. 

The situation changes drastically for the case of large $R, ~ R\gg
1/\sqrt{\sigma}$. Here one has
\be
E_n(R)=\sigma R 
+\frac{3}{2^{1/3}}\sigma^{1/3}\frac{(n+3/2)^{2/3}}{R^{1/3}},
\label{EL}
\ee
$$
\mu_n(R)=\frac{4\sigma^{1/3}(n+3/2)^{2/3}}{R^{1/3}},~~
\nu_{1,2 n}(R)=\frac{\sigma R}{2}.
$$
In this case $J_{1,2}=\frac{1}{6} \sigma R \gg\mu_n$, and the potential
regime becomes unadequate.

The case of large $R$ can be treated exactly in the einbein field method.
There are two different kinds of excitations, along the $Q \bar Q$ axis
and in the transverse direction, which are decoupled in the limit of
large $R$. In this case one has
\be
E_n(R)=\sigma R
+\frac{3}{2^{1/3}}\frac{\sigma^{1/3}(n_z+1/2)^{2/3}}{R^{1/3}}+
\frac{2\cdot 3^{1/2}}{R}(n_\rho+\Lambda+1),
\label{full}
\ee
where $\Lambda=\left\vert\frac{\vec L\vec R}{R}\right\vert$ is the
projection of orbital momentum onto $Q \bar Q$ axis ($z$ axis). Note,
that the subleading corrections due to the 
longitudinal and transverse vibrations are different,
the former behaves as $\frac{\sigma}{R}^{1/3}$, like the pure potential
regime (\ref{EL}), while the latter displays string-type behaviour
$\sim \Lambda/R$. 

The quasiclassical limit of large $\Lambda$, where only rotations 
around $z$ axis are taken into account, was found for the Hamiltonian
(\ref{Ham}) in \cite{vibr}. The large $R$ limit reads
\be
E(R) =\sigma R+2\sqrt{3}\frac{\Lambda}{R},
\label{Quasi}
\ee
which should be compared with the predictions from naive Nambu-Goto
string model
\be
E(R) =\sigma R+\frac{\pi\Lambda}{R}
\label{NG}
\ee  
in the small oscillation limit. The coefficients $\pi$ and $2\sqrt{3}$
are close to each other, and differ mainly due to the fact that string
configurations differ: there are two straight-line strings in the QCD
string model, and one continuous string in the Nambu-Goto case. The 
quasiclassical excitation curve \cite{vibr} is shown in Fig.2
together with the Nambu-Goto (\ref{NG}) and potential curves. Note
the absence of the unphysical divergent $1/R$ behaviour at small $R$ both
for quasiclassical and potential curves.

\begin{figure}[t!]
\epsfxsize=11.6cm
\centering
\epsfbox{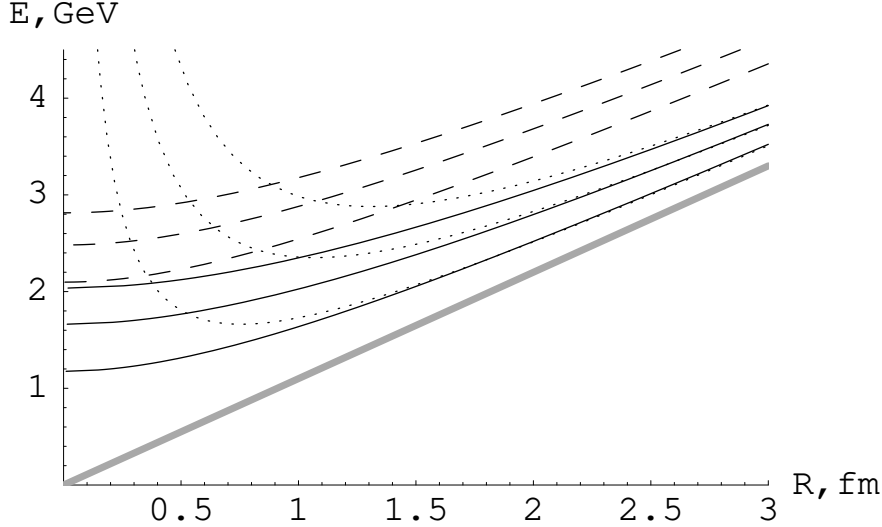}
\caption{Adiabatic hybrid potentials in various regimes. Quasiclassical
 (solid line), potential (dashed), and flux-tube
(dotted) curves for $n_z=n_\rho=0$ and $\Lambda=1,2,3$. The lowest curve is
$\sigma R$. $\sigma = 0.22$GeV$^2$.}
\end{figure}

\begin{figure}[t]
\epsfxsize=16.5cm
\centering
\epsfbox{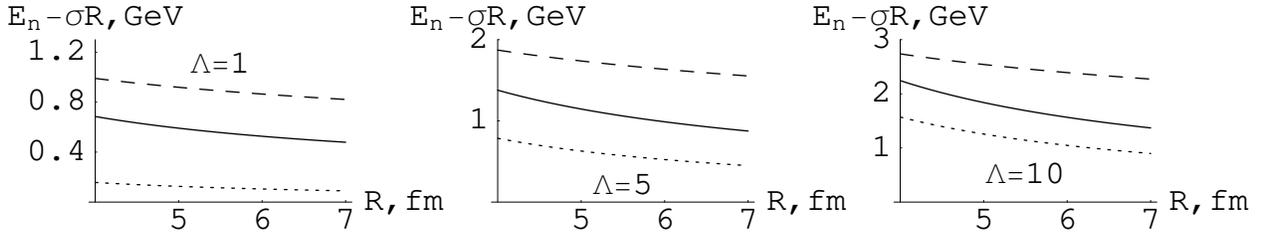}
\caption{Corrections to linear behaviour of potentials. QCD string (solid
line), potential (dashed), and flux-tube (dotted) curves; $n_z=n_{\rho}=0$;
$\sigma = 0.22$GeV$^2$.}
\end{figure}

The dominant subleading regime is defined by longitudinal motion: even if
no longitudinal quanta are excited, there is the contribution of zero
longitudinal oscillations in (\ref{full}). Still, if the distances are not
asymptotically large, the potential regime is substantionally contaminated 
by the string-type transverse vibrations.

Corresponding corrections to linear behaviour for QCD string (\ref{full}),
potential regime (\ref{EL}), and  Nambu-Goto string (\ref{NG}) are
shown in Fig.3.

 There is no QCD string calculations for large distances with
proper account of gluon spin yet, so the direct comparison with lattice
results is premature. Nevertheless, some preliminary conclusions can be
drawn. The behaviour (\ref{full}) displays the most pronounced difference
between QCD string and other models for constituent glue. In the flux tube
model \cite{fluxtube} the string vibrations are caused by string phonons, so
one expects the Nambu-Goto behaviour (\ref{NG}) at large distances. In
contrast to this, here the string vibrations are caused by pointlike
valence gluons. In the constituent gluon model with linear potential
\cite{const} one should have the potential-type behaviour (\ref{EL}),
while in the QCD string model the confining force follows from the minimal
area law, and, as a consequence, the contributions from the string inertia
leave room for the string-type vibrations.

\section{Full QCD string calculations in the potential regime} 

Let us now consider the regime of "small" $R$ relevant to the heavy hybrid
mass estimations. Actually these distances are not very small: one expects
that in the case of very heavy quarks the hybrid resides in the bottom
of the potential well given by the adiabatic curve. The lattice results
\cite{lattice} are not very accurate for small $R$, but the message is
quite clear: the bottom of the potential well is somewhere around 0.25 fm
for lowest curves. 

If only confining force is taken into account, the QCD string model
predicts the oscillator potential (\ref{ES}) with the minimum at $R=0$.
However, the minimum is shifted, if the long range confining force is
augmented by the short range Coulomb interaction, which is taken in the
form
\be
V_c = -\frac32\frac{\alpha_s}{r_1}-\frac32\frac{\alpha_s}{r_2}
+\frac{\alpha_s}{6R}.
\label{coul}
\ee
The coefficients in (\ref{coul}) are in accordance with the colour content
of the $Q \bar Q g$ system. Note that the $Q \bar Q$ Coulomb force is
repulsive. It is in contrast to the flux tube model \cite{fluxtube},
where the string phonons do not carry colour quantum number, so that the
$Q \bar Q$ pair is in the colour singlet state, and Coulomb interaction is
attractive. In the so-called single-bead version of the flux tube model
the minimum due the confining force is at $R=0$, and attractive 
Coulomb force cannot shift it, so the single-bead version seems to be
ruled out by lattice data. As the gluon energies of \cite{lattice} 
lie well below the curves (\ref{NG}), the simple Nambu-Goto regime 
is excluded too.  
 
Below we present the results of the QCD string model with Coulomb force
included, for small and intermediate values of $R$. The calculations were
performed in the potential regime, and the string inertia was taken into
account perturbatively, which is justified by arguments given above.

\begin{figure}[t!]
\epsfxsize=16.5cm
\centering
\epsfbox{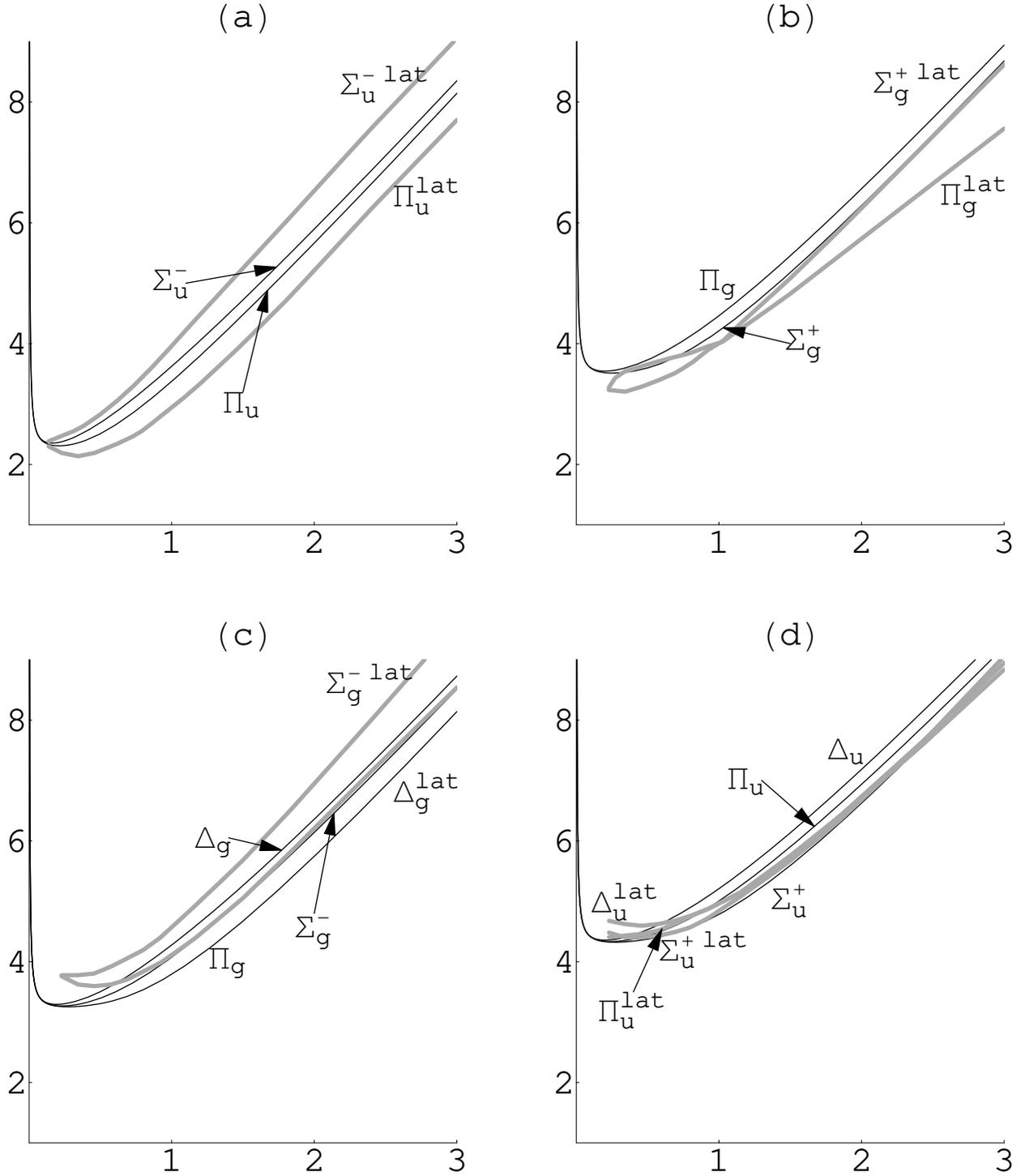}
\caption{Hybrid potentials in full QCD string model (thin solid
curves) compared to lattice results (thick light curves). $Q \bar Q$
distance $R$ is in units $2r_0\approx$ 1 fm and potentials $V$ are in units
$1/r_0$.}
\end{figure}

Varying over $\mu_i$ with result $\mu_{1,2}=\sigma r_{1,2}$, we obtain the
Hamiltonian in the form
\be
\tilde H=\frac{\mu}{2}+\frac{p^2}{2\mu}+\sigma(r_1+r_2)-\frac{3\alpha_s}{2}
\left(\frac{1}{r_1}+\frac{1}{r_2}\right)+\frac{\alpha_s}{6R}.
\label{H1}
\ee

We calculate unperturbed adiabatic energy levels variationally with
Gaussian wave functions used,
\be
E_{jl\Lambda}(\mu,R)=\langle \Psi_{jl\Lambda}(\vec r)|
\tilde H(\mu,\vec r,\vec R)|\Psi_{jl\Lambda}(\vec
r)\rangle\equiv \langle\tilde H\rangle_{jl\Lambda},
\label{E1}
\ee
\be
\Psi_{jl\Lambda}(\vec r)=\phi_l(\vec r)\sum_{\mu_1\mu_2}
C^{j\Lambda}_{l\mu_1 l\mu_2}Y_{l\mu_1}\left(\frac{\vec
r}{r}\right)s_{1\mu_2},
\label{Psi}
\ee
where $s$ is the spin wave function. The
unpertubed adiabatic potentials 
\be
V^0_{jl\Lambda}(R)=E_{jl\Lambda}(\mu^*(R), R),
\label{V0}
\ee
depend on the extremal value $\mu^*$ defined from the condition 
$\frac{\partial E_{jl\Lambda}(\mu, R)}{\partial\mu}=0$.

The string correction Hamiltonian  at $\nu\ll\mu$ after integrating out
$\nu$ reads
$$
H^{\mathrm{string}}=
-\frac{\sigma}{6\mu^2}\left(\frac{1}{r_1}L_1^2+\frac{1}{r_2}L_2^2 \right)=
$$
$$
-\frac{\sigma}{6\mu^2}\left\{\left(\frac{1}{r_1}+\frac{1}{r_2}\right)
\left[L^2+\frac14\left(R_jp_kR_jp_k-R_jp_kR_kp_j\right)\right]+\right.
$$
\be
\left.\frac12\left(\frac{1}{r_1}-\frac{1}{r_2}\right)\left[r_jp_kR_jp_k-r_jp_kR_kp
_j+R_jp_kr_jp_k-R_jp_kr_kp_j\right]\right\},
\label{Hstring}
\ee
where $\vec L_{1,2}=\vec r_{1,2}\times\vec p,~\vec L=\vec r\times\vec p$.

Here we estimate the contribution of the string correction (\ref{Hstring})
taking it at $R=0$, where it reads
\be
V_l^{\mathrm{string}}=-\frac{\sigma
l(l+1)}{3\mu_l^{*2}}\langle\frac{1}{r}\rangle_l.
\label{Vstring}
\ee
By the same way we consider spin-orbit correction at $R=0$,
\be
V^{\mathrm{SO}}_{jl}=-\frac{\sigma \vec S\vec L}{2\mu_l^{*2}}
\langle\frac{1}{r}\rangle_l,
\label{VSO}
\ee
where $ \vec S\vec L=\frac12(j(j+1)-l(l+1)-2)$.
We also subtract  a constant that corresponds to gluon self-energy. 
From comparison with lattice data
\cite{lattice} its value is found to be $\Sigma_{\mathrm{gl}}=473$ MeV.

In Fig.4 (a)-(d) potential
\be
V_{jl\Lambda}(R)=V^0_{jl\Lambda}(R)+V_l^{\mathrm{string}}+V^{\mathrm{SO}}_{j
l}- \Sigma_{\mathrm{gl}}
\label{V}
\ee
(solid curves) is compared to lattice potentials from \cite{lattice} (thick
grey curves). Parameters $\alpha_s=0.225,~\sigma=0.227$ GeV$^2$ are taken
from fit of the lattice Coulomb+linear ground-state potential.
In the Fig. 4 standard notations from the physics of atomic molecules are
used. 

In Table 1 the quantum numbers of levels, shown in Fig.4 (a)-(d), are
listed, in terms of $j,l,\Lambda$, and atomic notations. Note, that in the
QCD string model the gluon is effectively massive and has three
polarizations \cite{S}, and only two of them are excited with magnetic
components of field strength correlators, used in the lattice
calculations. It is just these states which are listed in the Table 1.
For more details justifying such correspondance see \cite{S}.

\begin{table}[t!]
\centering
\small
\caption{Quantum numbers of levels of Fig. 4}
\vspace{0.3cm}
\label{table}
\begin{tabular}{|c|c|c|}
  \hline
  (a) & $j=1,~l=1,~\Lambda=0,1$ & $\Sigma_u^-,~\Pi_u$ \\
  (b) & $j=1,~l=2,~\Lambda=0,1$ & $\Sigma_g^+,~\Pi_g$ \\
  (c) & $j=2,~l=2,~\Lambda=0,1,2$ & $\Sigma_g^-,~\Pi_g,~\Delta_g$ \\
  (d) & $j=2,~l=3,~\Lambda=0,1,2$ & $\Sigma_u^+,~\Pi_u,~\Delta_u$ \\
  \hline
  \end{tabular}
 \end{table}

\section{Conclusions}

The first results of the QCD string model for the hybrid adiabatic
potentials look rather encouraging. We have obtained the reasonable
agreement with lattice data at small and intermediate interquark
distances. 
The most interesting feature of the QCD string model is the large distance
behaviour of the adiabatic potentials, with potential-type longitudinal
and string-type transverse vibrations. The full calculations 
of large distance behaviour with proper account of gluonic spin are 
in progress now.  

\section{Acknowledgements}

Financial support of RFFI grants 00-02-17836, 00-15-96786, 
INTAS-RFFI grant IR-97-232 and INTAS CALL 2000-110 
is gratefully acknowledged.

\end{document}